\documentclass[man,floatsintext]{apa7}
\usepackage{graphics}
\usepackage[american]{babel}
\usepackage{float}
\usepackage{csquotes}
\usepackage[style=apa,sortcites=true,sorting=nyt,backend=biber]{biblatex}
\DeclareLanguageMapping{american}{american-apa}
\title{Exploring the relationship between surface features and explaining quality of YouTube explanatory videos}
\shorttitle{Explaining quality of YouTube explanatory videos}

\authorsnames[1,1,2,1,1,3]{Philipp Bitzenbauer, Sebastian Höfler, Joaquin M. Veith, Bianca Winkler, Tim Zenger, Christoph Kulgemeyer}
\authorsaffiliations{{Friedrich-Alexander-Universität Erlangen-Nürnberg, Professur für Didaktik der Physik, Staudtstr.~7, 91058 Erlangen, Germany},{Stiftungsuniversität Hildesheim, Institut für Mathematik und Angewandte Informatik, Samelsonplatz 1, 31141 Hildesheim, Germany},{University of Bremen, Institute for Science Education, Physics Education Department, Otto-Hahn-Allee 1, 28359 Bremen, Germany}}

\authornote{
\addORCIDlink{Philipp Bitzenbauer}{0000-0001-5493-291X}\\
\addORCIDlink{Joaquin M. Veith}{0000-0002-2147-0349}\\
\addORCIDlink{Christoph Kulgemeyer}{0000-0001-6659-8170}

Correspondence: philipp.bitzenbauer@fau.de
}

\abstract{Physics education research on explanatory videos has experienced a boost in recent years. Due to the vast number of explanatory videos available online, e.g. on YouTube, finding videos of high explaining quality is a challenging task for learners, teachers, and lecturers alike. Prior research on the explaining quality of explanatory videos on classical mechanics topics has uncovered that the surface features provided by YouTube (e.g. number of views or likes) do not seem to be suitable indicators of the videos' explaining quality. Instead, the number of content-related comments was found to be statistically significantly correlated with the explaining quality. To date, these findings have only been observed in the context of explanatory videos on classical mechanics topics. The question arises whether similar correlations between the explaining quality and YouTube surface features can be found for videos on topics that are difficult to access visually and verbally, for example from quantum physics. Therefore, we conducted an exploratory study analyzing the explaining quality of $N = 60$ YouTube videos on quantum entanglement and tunnelling. To this end, we made use of a category-based measure of explanatory videos' explaining quality from the literature. We report correlations between the videos' explaining quality, and the surface features provided by YouTube. On the one hand, our results substantiate earlier findings for mechanics topics. On other hand, partial correlations shed new light on the relationship between YouTube's surface features and explaining quality of explanatory videos.}


\begin{document}
\maketitle

\section{Introduction}
\label{sec1}
Quantum mechanics is a topical theme of physics in general \parencite[cf.][]{Acin2018}, and of physics education research in particular \parencite[cf.][]{BitzenbauerMDPI2021}. With today’s technological advancements, students may not only come into contact with quantum physics in formal learning settings, e.g., in undergraduate university courses (cf.~\cite{Singh2001,Marshman2019,Zhu2012,Zhu2012a,Passante2015,Pearson2010a,Galvez2005}), but also in the informal context: For example, interested learners can access the quantum world via multiple digital resources, such as smart-phone/tablet apps \parencite[e.g.,][]{Oss2015}, AR/VR applications \parencite[e.g.,][]{Suprapato2020,Dorland2019}, games \parencite[e.g.,][]{Chiofalo-2022,Zekir2022} or explanatory videos \parencite[e.g.,][]{Bitzenbauer2021b}.\\

Explanatory videos are brief videos - typically up to 10 minutes maximum - aimed at introducing and explaining a certain topic of interest \parencite[cf.][]{WolfKratzer2015neu}. They have increasingly been discussed in science education research in recent years \parencite[e.g.,][]{Pekag2010,Schroeder2017,Kulgemeyer2022}, both in the context of formal and informal learning environments, in particular on YouTube \parencite[e.g.,][]{Kulgemeyer2016neu,BeautempsBresges2021neu,Pattier-2021}. In the literature, factors that seem to be conducive to the success and popularity of explanatory YouTube videos on scientific topics have been revealed \parencite{BeautempsBresges2021neu,WelbourneGrant2016neu}, e.g., regarding the structure of a video \parencite{BeautempsBresges2021neu}. While it is desirable to reach as many people as possible, the main goal associated with the development of explanatory videos, of course, is to support student learning. Besides ensuring the success of the video, creators thus have to increase the quality of the explanations offered in their explanatory videos.\\

From the physics education research perspective, it is crucial to assist learners, teachers and university lecturers in selecting videos with high explanation quality from the plethora of (online) resources. In the case of YouTube explanatory videos, their popularity is publicly shown by means of different surface features, such as the number of views, the ratings of the video (e.g., the number of likes), or via the comment section. However, it remains open as to whether or not these surface features indeed correlate with the explanatory video's explaining quality, and hence, may serve as some kind of quality indicator in this respect. In other words: Can teachers and students rely on them?\\

This question has already been posed by \textcite{Kulgemeyer2016neu}. The authors presented a measure of explaining quality to investigate the above-mentioned question in the context of YouTube explanatory videos on two topics from classical mechanics, namely Newton's third law of motion and Kepler's laws \parencite{Kulgemeyer2016neu}. In their exploratory study, the number of content-related comments given by users below a specific video turned out to be the only variable that was statistically significantly correlated with the explaining quality of explanatory videos - neither the number of views, nor the number of likes or dislikes showed  correlations to explaining quality that were statistically significant \parencite{Kulgemeyer2016neu}. \textcite{Kulgemeyer2016neu} see the need for further studies on the relationship between surface features provided by YouTube and explaining quality, in particular, regarding other topics. They developed a hypothesis on this relationship that requires further evidence. Videos on topics from quantum physics seems to add a valuable perspective here.\\

Quantum physics differs fundamentally from classical mechanics, especially since its concepts are not directly visible with the naked eye. Thus, explanations of quantum physics topics arguably require specifically varied explanations. As a result, the question arises as to whether or not the metrics of YouTube explanatory videos about quantum concepts show similar correlations to an established measure of explaining quality as has previously been revealed by \textcite{Kulgemeyer2016neu} for explanatory videos on classical mechanics topics. This is where this research project comes in: We investigate the explaining quality of YouTube explanatory videos on two genuine quantum physics topics without classical analogies, namely quantum entanglement and quantum tunnelling. To this end, the research methods used by \textcite{Kulgemeyer2016neu} were leveraged into our study. The objective of the research project presented in this article is to expand on Kulgemeyer and Peters' results by exploring correlations between the YouTube surface metrics (e.g., likes, dislikes, views, number of days since release, number of relevant comments) of explanatory videos on these two quantum topics and the explaining quality of these videos \parencite[cf.][]{Kulgemeyer2016neu}.

\section{Research Questions}
\label{sec2}
The present study addresses the following research questions:
\begin{enumerate}
    \item[1.] How is the explaining quality of YouTube explanatory videos on quantum entanglement and quantum tunnelling correlated with the videos' metrics such as the number of views, the number of likes, or the number of dislikes?
    \item[2.] How is the number of content-related comments correlated with the explaining quality of YouTube explanatory videos on quantum entanglement and quantum tunnelling? 
\end{enumerate}


\section{Research Background}
\label{sec3}
\subsection{Explaining physics}
\label{sec:3_1}
Instructional explanations are ``designed with the specific purpose of teaching a student or group of students'' \parencite[][p. 90]{LeinhardtSteele2005neu}. Hence, instructional explanations need to be distinguished from scientific explanations \parencite[][]{Treagust1999}. \Textcite{WittwerRenkl2008neu} uncovered four factors that lead to effective instructional explanations; for example, they should...
\begin{enumerate}
    \item ...``be adapted to the learner's knowledge prerequisites'' \parencite[][p. 51]{WittwerRenkl2008neu}, 
    \item ...``focus on concepts and principles'' \parencite[][p. 53]{WittwerRenkl2008neu},
    \item ...``should be integrated into the learners' ongoing cognitive activities'' \parencite[][p. 55]{WittwerRenkl2008neu}, and 
    \item ...``should not replace learners' knowledge-construction activities'' \parencite[][p. 56]{WittwerRenkl2008neu}.
\end{enumerate} 

These factors have been expanded to a total of nine factors in a 2019 review addressing instructional explanations in science teaching \parencite[][p. 90]{Kulgemeyer2019-Stud}. An important criterion for effective instructional explanations is the \textit{adaption to the explainee} because this criterion mirrors that explaining is to be regarded a constructivist process \parencite[][]{Kulgemeyer2016neu}.\\

The constructivist nature of explanations is reflected in the communication model for explaining physics presented by \textcite{KulgemeyerSchecker2013neu}. This model consists of four pillars, namely the explainer, the explanation itself, the explainee, and the explainee's feedback. The fact that a good explanation requires 
\begin{enumerate}
    \item constant evaluation of the explainee's feedback, and 
    \item prompt adaptation of the explanation based on that feedback,
\end{enumerate}
is at the heart of this model \parencite[][]{KulgemeyerSchecker2013neu}. According to the communication model for explaining physics, ``the explainer can vary the explanation on four levels based on this feedback, ranging from the language code, the graphic representation form and the mathematic code, to using examples and analogies'' \parencite[][p. 3]{Kulgemeyer2016neu}.

\subsection{Design principles for explanatory videos} 
The \textit{Cognitive load theory} \parencite[e.g.,][]{Sweller1988Neu,Sweller1994neu,Sweller1998Neu} assumes a limited capacity of working memory caused by a cognitive load on learners in learning environments, which - in its modern view \parencite[cf.][]{Sweller-2019} - is composed by 
\begin{itemize}
    \item \textit{intrinsic cognitive load} which is dependent on the concrete learning task, the students' prior knowledge, or the teaching materials used, and
    \item \textit{extraneous cognitive load} stemming from irrelevant cognitive processes that tie up working memory capacities and thus hinder the learning process.
\end{itemize}
According to \textcite{Sweller-2019}, the Cognitive load theory ``provides evidence-informed principles that can be applied to the design of instructional messages or relatively short instructional units, such as lessons, written materials consisting of text and pictures, and educational multimedia'' (p. 274).\\

The \textit{Cognitive Theory of Multimedia Learning} \parencite[cf.][]{Mayer-1999} builds upon the above-mentioned Cognitive load theory. This theory is based on three fundamental assumptions that, taken together, describe how auditory-verbal or visual-imagery information is processed toward long-term memory: 
\begin{itemize} 
    \item The Dual-Channel assumption describes that ``humans possess separate channels for processing visual and auditory information'' \parencite[][p. 63]{Mayer-2009}.
    \item The Limited-Capacity assumption describes that each of the above-mentioned channels can only process  a limited amount of ``chunks'' \parencite[][p. 67]{Mayer-2009} of information simultaneously. 
    \item The Active-Processing Assumption describes that students' active engangement is necessary for students constructing knowledge \parencite[][]{Mayer-2009}. 
\end{itemize}
Both the Cognitive load theory and the Cognitive Theory of Multimedia Learning have been the basis for prior research on explanatory videos aimed at fostering student learning \parencite[cf.][]{Noor-2014,Kruger-2016}. In addition, different studies have derived design principles that may influence the effectiveness of explanatory videos against the backdrop of the above-mentioned theories \parencite[e.g.,][]{Brame2016neu,Muller-2008,Kay2014neu}. For example, it has been indicated that the integration of interactive elements into explanatory videos \parencite[][]{Delen2014} or the use of a 1st person perspective in explanatory videos \parencite[][]{Fiorella2017} might have a positive impact on students' performance. \Textcite{Findeisen2019} reviewed and systematized studies dealing with potential effects of explanatory videos' design principles on student learning, and derived guidelines for the development of explanatory videos based on the overall picture emerging from current empirical findings. 

\subsection{Explaining quality of explanatory videos} 
\label{sec:3_3}
In the previous sections, we reviewed both the current state of research on explaining physics and on design criteria for the development of explanatory videos. In this section, both perspectives are merged in order to shed more light on the state of research on the explanatory quality of explanatory videos.  \\

\Textcite{Kulgemeyer2020Framework} presented a framework for effective explanation videos. This framework is  
\begin{itemize} 
    \item ...consistent with guidelines on the quality of explanatory videos published elsewhere in the literature \parencite[e.g.,][]{Findeisen2019,Brame2016neu}, and
    \item ...acknowledges research on multimedia learning \parencite[][]{Kulgemeyer2020Framework},
\end{itemize} 
while building upon state-of-the-art research on instructional explanations \parencite[e.g.,][]{Geelan2012,WittwerRenkl2008neu}. In this framework, seven factors comprising a total of 14 features are described to have an impact on the effectiveness of explanatory videos \parencite[][p. 2450]{Kulgemeyer2020Framework}. Examples are the use of summaries (factor: structure of the video), the use of an appropriate language-level (factor: tools for adaption), the avoidance of digressions (factor: minimal explanation), or the adaption to prior knowledge, misconceptions and interest (factor: adaption). An overview of the whole framework for effective explanation videos is presented in \textcite[][p. 2450]{Kulgemeyer2020Framework}.\\

The above-mentioned framework has been tested empirically in order to clarify as to whether or not an explanatory video developed with respect to the framework leads to higher student achievement compared to a video that has not strictly been developed according to the framework \parencite[][]{Kulgemeyer2020Framework}. The results of this study revealed that students learning with an explanation video adhering strongly to this framework showed significantly more declarative knowledge in a post-test than students learning with a video that has not strictly been developed according to the framework ($d = 0.42$). However, no statistically significant difference in the post-test scores regarding conceptual knowledge was observed.

\subsection{\label{sec:34} Evaluation of explanatory videos' explaining quality}

An online test which allows for the assessment of physics explanatory skills has been published by \textcite{Bartels-2019}. This test has been developed both for its usage in teacher education and for self-assessment.\\

Moreover, based on the communication model for explaining physics \parencite[][]{KulgemeyerSchecker2013neu}, \textcite[][p. 121]{KulgemeyerTomczyszyn2015neu} developed a process-oriented and category-based measure for the assessment of explanation skills. \Textcite[][]{Kulgemeyer2016neu}, adopted this category-based measure for the evaluation of explanatory videos' explaining quality. The  category system to evaluate explanatory videos' explaining quality (cf. appendix) consists of seven main categories (content, structure, use of language, contexts and examples, mathematics, interrogation, non-verbal elements) comprising a total of 31 different categories. Each of these categories is either assigned to a certain explanatory video (= 1 point) or not (= 0 points). Four out of the 31 subcategories (1. scientific mistake, 2. ignoring students' comment, 3. leaving new technical term uncommented, 4. without context) are related to a decrease of explaining quality, and hence, a negative point (= -1 point) is allocated to the video for their occurrence.\\

Within the scope of evaluating the explaining quality of explanatory videos (i.e., in the course of categorisation), each category is considered uniformly and there is no counting of a successive occurrence of the same category, ``since repetitions of the same wording or the repeated use of a similar explaining aid without any variation are not considered a rich and varied explanation'' \parencite[][p. 6]{Kulgemeyer2016neu}. By summing up the points received on the basis of the categories assigned, a specific number of ``category points'' \parencite[][p. 6]{Kulgemeyer2016neu}, referred to as CP, can be calculated for a given explanatory video \parencite[][p. 6]{Kulgemeyer2016neu}: $$\textrm{CP} = \sum X_{+} + \sum X_{-},$$
where $X_{+}$ denotes the number of positive categories assigned to a video, and $X_{-}$ stands for the number of all negative categories assigned to a video. The category points (with the upper limit of $28$ CP) serve as a measure of an explanatory video's explaining quality as has been shown by \textcite{Kulgemeyer2016neu}. \\

It is important to note that the category points assigned to a specific explanatory video may neither judge the video's overall quality (e.g., a video's technical design is not taken into account), nor do the CP help finding the best explanation of a specific topic under investigation among multiple explanatory videos. Instead, the rationale underlying this measure is ``to distinguish between rich and varied explanations on the one hand and those with fewer variations on the other'' because ``those with fewer variations in their explanations may be less suitable for a wider range of
viewers as some learners’ needs may not be considered'' \parencite[][p. 9]{Kulgemeyer2016neu}. 

\section{Methods} 
\label{sec4}
In this section, we outline the methodology applied in our exploratory study to approach a clarification of the research questions. We aim at expanding on Kulgemeyer and Peters's study according to which none of the correlations between the surface features provided for YouTube explanatory videos and their explaining quality was statistically significant, except from the number of content-related comments \parencite[][]{Kulgemeyer2016neu}. In a further study, \textcite{Kocyigit-2019} even conclude that the ``number of views, likes, dislikes, and comments per day is not a predictor of high-quality videos on YouTube'' (p. 1267).

\subsection{Sample}
\subsubsection{Content domain}
We decided to analyze YouTube explanatory videos on two topics: (a) quantum entanglement, and (b) quantum tunnelling. We analyzed videos addressing these topics because neither quantum entanglement nor quantum tunnelling has any classical analogy and the quantum physics formalism does not enable a space-time description of these concepts \parencite[cf.][]{Ubben2022}. In this way, our study allows best to contrast the previous findings of \textcite{Kulgemeyer2016neu} who analyzed explanatory videos on topics of classical mechanics.

\subsubsection{Inclusion-exclusion criteria and search procedure} 
Following \textcite{Kulgemeyer2016neu}, we found the videos to be included in our sample via YouTube's search engine applying the search strings ``quantum entanglement'' and ``quantum tunnelling'', respectively. We used the following inclusion-exclusion criteria for selecting videos appropriate for data analysis: 
\begin{itemize}
    \item The video is published in the English language. 
    \item The video exclusively covers one of the two topics quantum entanglement or quantum tunnelling, respectively. Videos that covered both topics were excluded.  
    \item Video-recorded lectures (or excerpts thereof) were excluded, since recorded lectures ``do not share the explainers’ intentional core of publishing a concise explanatory video'' \parencite[][p. 4]{Kulgemeyer2016neu}. 
    \item The video has a maximum duration of 10 minutes.
    \item The video is scientifically sound \parencite[cf.][]{Kulgemeyer2016neu}.
\end{itemize}
The latter criterion was important because it only makes sense to compare ``the explaining quality of scientifically correct explanations'' \parencite[][p. 5]{Kulgemeyer2016neu}. Applying the above-mentioned search strings, we found more than 100.000 videos on both topics. A title-caption screening of the search results led to the exclusion of the majority of these videos since they did not fulfill the inclusion criteria (in this stage most often due to a duration above 10 minutes, the coverage of topics beyond the ones under investigation, or representing recorded lectures). In a next step, we reviewed about 200 videos on each of the topics quantum entanglement or tunnelling in detail. Again, we excluded the videos that did not fulfill the inclusion criteria (in this stage most often due to serious scientific errors). Lastly, for our final sample, we (a) settled on videos with comparable run-times of around 5 minutes as has been done in the prior study conducted by\textcite[][]{Kulgemeyer2016neu}, and (b) aimed for a sample size comparable to the one of the earlier study in the classical mechanics context \parencite[][]{Kulgemeyer2016neu}. The final sample consists of $60$ YouTube explanatory videos that were included for data analysis, $30$ of which address the topic of quantum entanglement, and $30$ of which focus on quantum tunnelling. 

\subsubsection{Description of the sample}
The mean duration of the selected videos is $m = 4.97$ min with a standard deviation of SD $= 2.43$ min. The explanatory videos on quantum entanglement ($m = 4.74$ min, SD $= 2.38$ min) were of similar length as those on quantum tunnelling ($m = 5.20$ min, SD $= 2.48$ min). Moreover, the videos in our sample are of similar length as the ones included in the prior study \parencite[cf.][]{Kulgemeyer2016neu}.

\subsection{Data collection} 
The explanatory videos included in the final selection have been analyzed in August and September 2021. For the exploration of our research questions, the data collection comprised three aspects: In a first step, we collected each videos' surface features, i.e., the number of \textit{likes} and \textit{dislikes}, the number of \textit{views}, and the \textit{publication date} to calculate the videos' \textit{time online} (in days). Additionally, we recorded the \textit{number of subscribers} to the channels by which the videos were published. The \textit{average view duration} has been a further surface feature which was included in Kulgemeyer and Peter's study on explanatory videos on classical mechanics topics \parencite[][p. 5]{Kulgemeyer2016neu}. However, at the time of conducting our data collection this feature was not publicly accessible anymore and hence, it is not included in our analysis. In addition, the dislike statistic is not publicly available anymore since the end of 2021 - since our data collection was conducted in August and September 2021, however, we kept the number of dislikes for each video in our dataset and also included it in the data analysis. This allows for a more comprehensive comparison to the earlier results published by \textcite{Kulgemeyer2016neu} and may help to better understand the interaction of users with explanatory videos. For a description of all the above-mentioned YouTube metrics, we refer the reader to the \textcite[][]{YouTube}. \\

In a second step, we categorised the comments given below the videos in order to receive the number of relevant comments for each video. We provide a proper description of (a) the term \textit{relevant comment} and (b) the categorisation procedure in the data analysis section. We explored relevant comments because they ``provide by far the most intense communication channel between explainer and addressee'' \parencite[][p. 5]{Kulgemeyer2016neu}.\\

Lastly, following the data collection method from \textcite{Kulgemeyer2016neu}, we used the category system described above (cf. appendix) to assess the explaining quality of the explanatory videos included in our sample. The coding was performed by two independent raters. The inter-rater reliability expressed via Cohen's Kappa can be regarded \textit{substantial} ($\kappa = 0.79$) according to \textcite{Cohen1988}. Against this backdrop, the category system used in this study allows for an objective assessment of the explaining quality of explanatory videos. Furthermore, the reliability of the measure has been found to be satisfactory (Cronbach's $\alpha = 0.58$; in the earlier study by \textcite{Kulgemeyer2016neu}, a comparable value of $\alpha = 0.69$ has been reported). Moreover, the category-system used for this study allows for a valid measure of explanatory videos' explaining quality as has been justified by \textcite{Kulgemeyer2016neu}.   \\

As a last step of data collection, we calculated the category points CP for each explanatory video included in our sample. These category points were then further processed to data analysis. 

\subsection{Data analysis carried out the answer research question 1}
We report descriptive statistics (range, median Mdn, mean $m$, standard deviation SD) regarding the category points of the explanatory videos on quantum entanglement and quantum tunnelling, respectively. \\

We conducted a correlation analysis in order to explore relationships between the videos' explaining quality (in category points CP) on the one hand, and the surface features provided by YouTube on the other hand. We report Pearson's correlation coefficient $r$ because the data are of metric scale. We interpret correlation coefficients according to \textcite{Cohen1988}: weak correlation for $0.1 < |r| < 0.3$, moderate correlation for $0.3 \leq |r| < 0.5$, strong correlation for $|r| \geq 0.5$. In addition, we report partial correlations to verify that observed relationships are no artefact caused by 
\begin{itemize}
    \item the videos' time online, i.e. the time that has passed between the publication of a video and the data collection, and
    \item the number of subscribers to the channels by which the videos were published.
\end{itemize}
The latter control variable seems particularly important due to the fact that the YouTube algorithms promote videos published by popular channels which in turn leads to high numbers of views for these videos. This might influence the results, and hence, deserves special attention.

\subsection{Data analysis carried out the answer research question 2}
\label{analysis_RQ2}
The comments below each video included in our sample have been categorised. For the categorisation, we used the category system presented by \textcite[][p. 8]{Kulgemeyer2016neu} which consists of four categories: 
\begin{enumerate}
    \item Comment on content: ``further question or comment on notations'' \parencite[][p. 8]{Kulgemeyer2016neu}.
    \item Comment on explanation: ``constructive criticisms and inquiries for more videos'' \parencite[][p. 8]{Kulgemeyer2016neu}.
    \item Comment on explainer's style: ``comments on the style including a reason'' \parencite[][p. 8]{Kulgemeyer2016neu}.  
    \item Comment on use: description of ``the viewer's use of the video, e.g., revising, preparing a talk or learning for a test'' \parencite[][p. 8]{Kulgemeyer2016neu}.
\end{enumerate}
All comments that could be assigned to at least one of these categories, were considered as \textit{relevant comments}. Comments that could not be assigned to any of these categories, conversely, were excluded from further analysis because they were not related specifically to the content presented in the respective video or to the explanation offered within. For the further analysis, we refrained from a deeper differentiation between the different categories as has been done by \textcite{Kulgemeyer2016neu} because research question 2 only addresses relevant comments in general. \\

The categorisation of the all comments underneath $N = 60$ explanatory videos included in our sample led to a total of $1452$ relevant comments. The number of relevant comments for each video was included in our data set as a metric variable and was used for correlation analysis. Again, we additionally calculated partial correlations to verify that observed relationships are no artefact caused by the videos' time online, or the number of subscribers to the channels by which the videos were published.

\section{Results} 
\label{sec5}

\subsection{Descriptives} 
The median value of the explanatory videos' explaining quality (measured in CP) was Mdn $= 11$ CP for our total sample, ranging from $2$ CP (one video) to $18$ CP (two videos). In table~\ref{tab:descr}, descriptive statistics on the category points assigned to the videos comprised in our sample are reported separately for the two subject areas under investigation, namely quantum entanglement and quantum tunnelling, respectively.

{\linespread{1}

\begin{table}
\caption{Descriptive statistics on the measure of explaining quality of the videos included in our sample (expressed in category points CP).}
\label{tab:descr} 
\begin{tabular}{p{0.4\textwidth}p{0.1\textwidth}p{0.1\textwidth}p{0.1\textwidth}p{0.1\textwidth}}
\toprule
   & Range & Mdn & $m$ & SD   \\
\midrule
Explaining quality \newline Total sample~\tabfnm{a} & $2-18$ & $11.00$ & $11.02$ & $3.28$ \\
\midrule
Explaining quality \newline quantum entanglement~\tabfnm{b} & $2-18$ & $11.00$ & $11.03$ & $3.62$\\
\midrule
Explaining quality \newline quantum tunnelling~\tabfnm{c} & $4-18$ & $11.50$ & $11.00$ & $2.96$ \\
\bottomrule
\end{tabular}
{\linespread{1.5}
\begin{tablenotes}[para,flushleft]
        {\small
            \textit{Note.} 
            \tabfnt{a} $N = 60$.
            \tabfnt{b} $N = 30$.
            \tabfnt{c} $N = 30$.
            
         }
    \end{tablenotes}}

\end{table}
}
\subsection{Correlation analysis} 
The correlation analysis results are summarized in table~\ref{tab:corr_overall}. Within the total sample, we find statistically significant correlations between the videos' explaining quality and the number of views ($r = 0.27$, $p < 0.05$), the number of likes ($r = 0.37$, $p < 0.01$), and the number of dislikes ($r = 0.32, p < 0.05$). The highest correlation is uncovered between the videos' explaining quality and the number of relevant comments ($r = 0.46$, $p < 0.01$), whereas the correlation between the videos' explaining quality and their time online does not differ from $0$ with statistical significance.

\clearpage

{\linespread{1}
\begin{table}[H]
\caption{Pearson's correlation coefficient $r$ between the measure of explaining quality (in CP) and the surface features (incl. number of relevant comments) for the total sample, the videos on quantum entanglement and the ones on quantum tunnelling, respectively. For all correlations, we report $95\%$ confidence intervals ($95\%$-CI).}
\label{tab:corr_overall}
\begin{tabular}{p{0.35\textwidth}p{0.25\textwidth}p{0.1\textwidth}p{0.2\textwidth}}
\toprule
                       & Surface feature & $r$ & $95\%$-CI   \\
\midrule
Explaining quality      & Time online (days) & $-0.14$ & $[-0.38; 0.12]$ \\
Total sample~\tabfnm{a}            & Views              & $0.27^{*}$  & $[0.02; 0.49]$\\
               & Likes              & $0.37^{**}$ & $[0.13; 0.57]$\\
                        & Dislikes            & $0.32^{*}$  & $[0.06; 0.53]$\\
                        & Relevant comments  & $0.46^{**}$ & $[0.24; 0.64]$\\
\midrule
Explaining quality     & Time online (days) & $0.00$      & $[-0.36; 0.36]$ \\
quantum entanglement~\tabfnm{b}    & Views              & $0.32$      & $[-0.05; 0.61]$\\
              & Likes              & $0.42^{*}$  & $[0.08; 0.68]$\\
                        & Dislikes           & $0.37$      & $[0.00; 0.65]$\\
                        & Relevant comments  & $0.59^{**}$ & $[0.29; 0.79]$\\
\midrule 
Explaining quality      & Time online (days) & $-0.28$     & $[-0.58; 0.09]$ \\
quantum tunnelling~\tabfnm{c}      & Views              & $0.23$      & $[-0.15; 0.54]$\\
              & Likes              & $0.30$      & $[-0.07; 0.60]$\\
                        & Dislikes           & $0.30$      & $[-0.07; 0.60]$\\
                        & Relevant comments  & $0.31^{(*)}$& $[-0.06; 0.60]$\\
\bottomrule
\end{tabular}
{\linespread{1.5}
\begin{tablenotes}[para,flushleft]
        {\small
            \textit{Note.} Statistical significance of the correlations is denoted by an asterisk:
            \tabfnt{(*)}\textit{p} < .10.
            \tabfnt{*}\textit{p} < .05.
            \tabfnt{**}\textit{p} < .01.
            \tabfnt{a} $N = 60$.
            \tabfnt{b} $N = 30$.
            \tabfnt{c} $N = 30$.
         }
    \end{tablenotes}}
\end{table}
}

A striking observation is the positive correlation between the number of dislikes and the measure of explaining quality, both in the total sample ($r = 0.37$, $p < 0.01$) and the two sub-samples including videos on quantum entanglement ($r = 0.37$) and quantum tunnelling ($r = 0.30$). In order to better understand the underlying principles, we decided to introduce three further variables into our analysis: 
\begin{enumerate}
    \item We calculated the \textit{likes-to-dislikes ratio} for each video included in our sample. This variable allows to contrast the frequency of occurrence of likes to that of dislikes for a given video, and thus could potentially be a more accurate measure of the quality of an explainer video. A similar approach has already been taken by~\textcite{Meyer2019}.
    \item We assumed that the interaction with a specific explanatory video, i.e., giving a like or a dislike to a video, requires the user to be cognitively activated to some extent. We therefore introduced the variable \textit{interactions} calculated via $$\textrm{interactions} = \sum \textrm{likes} + \sum \textrm{dislikes},$$
    to explore the relationship between explaining quality and the number of interactions. This might provide further insights into how users interact with explanatory videos depending on their explaining quality. 
    \item Lastly, to check as to whether the number of likes and dislikes are really relevant variables with respect to explanatory videos' explaining quality, we calculated the the \textit{likes-to-interactions ratio} via $$\frac{\textrm{likes}}{\textrm{interactions}},$$ because for a high quality explanatory video one could expect a high number of likes compared to the total number of interactions. Note that the \textit{dislike-to-interactions ratio} does not contain any further information, and hence, leads to mathematically equivalent results in the correlation analysis (up to sign). 
\end{enumerate}

The correlations between the measure of explanatory videos' explaining quality and the three above-mentioned variables are shown in table~\ref{tab:interaction}: It becomes apparent that neither the likes-to-dislike-ratio ($r = 0.11$) nor the likes-to-interactions ratio ($r = -0.03$) is correlated statistically significantly with the explaining quality of the explanatory videos included in our sample. The only variable showing moderate but statistically significant correlation to the videos' explaining quality is the number of interactions itself ($r = 0.39$, $p < 0.01$). From these results, of course, no particular (and a fortiori no causal) relationship between 
\begin{enumerate}
    \item the number of likes or dislikes (and their ratio), and 
    \item the videos' explaining quality can be inferred.
\end{enumerate}
\clearpage

{\linespread{1}
\begin{table}[H]
\caption{Pearson's correlation coefficient $r$ of the measure of explaining quality (in CP) with the variables \textit{likes-to-dislike ratio}, \textit{interactions}, and \textit{likes-to-interactions ratio}, respectively. For all correlations, we report $95\%$ confidence intervals ($95\%$-CI).}
\label{tab:interaction} 

\begin{tabular}{p{0.3\textwidth}p{0.35\textwidth}p{0.1\textwidth}p{0.15\textwidth}}
\toprule
                         & Variable & $r$ & $95\%$-CI   \\
\midrule
Explaining quality       & Likes-to-dislikes ratio     & $0.11$       & $[-0.21; 0.40]$ \\
Total sample~\tabfnm{a}            & Interactions                & $0.39^{**}$  & $[0.14; 0.59]$\\
                         & Likes-to-interactions ratio & $-0.03$      & $[-0.28; 0.23]$\\
\midrule
Explaining quality      & Likes-to-dislikes ratio     & $-0.01$       & $[-0.43; 0.41]$ \\
quantum entanglement~\tabfnm{b}     & Interactions                & $0.44^{*}$  & $[0.09; 0.69]$\\
                        & Likes-to-interactions ratio & $-0.03$      & $[-0.40; 0.34]$\\
\midrule 
Explaining quality     & Likes-to-dislikes ratio     & $0.22$       & $[-0.26; 0.61]$ \\
quantum tunnelling~\tabfnm{c}     & Interactions                & $0.31^{(*)}$  & $[-0.06; 0.61]$\\
                       & Likes-to-interactions ratio & $-0.03$      & $[-0.39; 0.34]$\\
\bottomrule
\end{tabular}
{\linespread{1.5}
\begin{tablenotes}[para,flushleft]
        {\small
            \textit{Note.} Statistical significance of the correlations is denoted by an asterisk:
            \tabfnt{(*)}\textit{p} < .10.
            \tabfnt{*}\textit{p} < .05.
            \tabfnt{**}\textit{p} < .01.
            \tabfnt{a} $N = 60$.
            \tabfnt{b} $N = 30$.
            \tabfnt{c} $N = 30$.
         }
    \end{tablenotes}
    }
\end{table} 
}

However, it seems that the total number of user interactions is correlated significantly with the videos' explaining quality - no matter of whether these interactions result in a like or a dislike in the end. It is necessary to control the correlations presented in tables~\ref{tab:corr_overall} and \ref{tab:interaction} for the videos' time online (in days), and the number of subscribers to the channels by which the videos were published in order to explore this in more detail. Therefore, we report partial correlations in the next subsection.

\subsection{Partial correlations} 
In this subsection, we report partial correlations which refer to the entire sample. This means that we do not distinguish between the sub-samples here for the sake of clarity.\\

Controlling the correlations between our explanatory videos' explaining quality (measured in CPs) and the YouTube surface features for the videos' times online, we observe the following (cf. table~\ref{tab:partial_time}): Besides significant correlations between explaining quality and the number of views ($r = 0.33$, $p < 0.05$), the number of likes ($r = 0.40$, $p < 0.01$), the number of dislikes ($r = 0.33$, $p < 0.05$), and the number of relevant comments ($r = 0.55$, $p < 0.001$), only the number of interactions ($r = 0.41$, $p < 0.01$) shows a significant correlation to the explaining quality of our videos. These partial correlations uncover similar relationships between YouTube's surface metrics and the videos' explaining quality as the ones presented earlier (cf. table~\ref{tab:corr_overall}).

{\linespread{1}
\begin{table}[H]
\caption{Partial correlations (controlled for the time online) between the measure of explaining quality (in CP) and YouTube's surface metrics as well as the \textit{likes-to-dislike ratio}, the number of \textit{interactions}, and \textit{likes-to-interactions ratio}.}
\label{tab:partial_time} 
\begin{tabular}{p{0.4\textwidth}p{0.35\textwidth}p{0.1\textwidth}}
\toprule
Controlled for: time online (days) & Variable & $r$   \\
\midrule
Explaining quality       & Views                      & $0.33^{*}$        \\
Total sample~\tabfnm{a}  & Likes                       & $0.40^{**}$ \\
                        & Dislikes                    & $0.33^{*}$     \\
                        & Relevant comments           & $0.55^{***}$        \\
                        & Likes-to-dislikes ratio     & $0.08$ \\
                        & Interactions                & $0.41^{**}$     \\
                        & Likes-to-interactions ratio & $-0.05$\\
\bottomrule
\end{tabular}
{\linespread{1.5}
\begin{tablenotes}[para,flushleft]
        {\small
            \textit{Note.} Statistical significance of the correlations is denoted by an asterisk:
            \tabfnt{*}\textit{p} < .05.
            \tabfnt{**}\textit{p} < .01.
            \tabfnt{***}\textit{p} < .001.
            \tabfnt{a} $N = 60$.
         }
\end{tablenotes}}
\end{table}
}

In a next step, we controlled for the number of subscribers to the channels by which the videos were published. The corresponding partial correlations are shown in table~\ref{tab:partial_subscribe}: Only three of the correlations remain statistically significant in this case, namely the ones between the explanatory videos' explaining quality and the number of likes ($r = 0.43$, $p < 0.01$), the number of relevant comments ($r = 0.47$, $p < 0.001$), and the number of interactions ($r = 0.43$, $p < 0.01$). In contrast, both the correlations of the videos' explaining quality to the number of views, and the number of dislikes are not statistically significant anymore. We will discuss these observations in the discussion section.

{\linespread{1}
\begin{table}[]
\caption{Partial correlations (controlled for the number of subscribers) between the measure of explaining quality (in CP) and YouTube's surface metrics as well as the \textit{likes-to-dislike ratio}, the number of \textit{interactions}, and \textit{likes-to-interactions ratio}.}
\label{tab:partial_subscribe} 

\begin{tabular}{p{0.4\textwidth}p{0.35\textwidth}p{0.1\textwidth}}
\toprule
Controlled for: \newline Number of subscribers & Variable & $r$   \\
\midrule
Explaining quality       & Views                      & $0.23$        \\
Total sample~\tabfnm{a}  & Likes                       & $0.43^{**}$ \\
                        & Dislikes                    & $0.26$     \\
                        & Relevant comments           & $0.47^{***}$        \\
                        & Likes-to-dislikes ratio     & $0.08$ \\
                        & Interactions                & $0.43^{**}$     \\
                        & Likes-to-interactions ratio & $-0.06$\\
\bottomrule
\end{tabular}
{\linespread{1.5}
\begin{tablenotes}[para,flushleft]
        {\small
            \textit{Note.} Statistical significance of the correlations is denoted by an asterisk:
             \tabfnt{*}\textit{p} < .05.
            \tabfnt{**}\textit{p} < .01.
            \tabfnt{***}\textit{p} < .001.
            \tabfnt{a} $N = 60$.
         }
\end{tablenotes}}
\end{table}}

\section{Discussion} 
\label{sec6}
In our exploratory study, we investigated as to how the explaining quality of YouTube explanatory videos on genuine quantum topics such as quantum entanglement and quantum tunnelling is correlated with the surface features provided by YouTube alongside each online video. In this section, we discuss the results of our study with regards to our research questions, and against the backdrop of a study published earlier that explored similar questions for explanatory videos on classical mechanics topics~\parencite[cf.][]{Kulgemeyer2016neu}.

\subsection{Discussion of research question 1}
\label{sec:disRQ1}
A correlation analysis revealed statistically significant correlations between the explanatory videos' explaining quality and the surface features provided by YouTube (cf. table~\ref{tab:corr_overall}): 
\begin{itemize} 
    \item The correlation between the number of views and the explanatory videos' explaining quality is small and statistically significant for the total sample ($r = 0.27$, $p < 0.05$) but not statistically significant for the videos on quantum entanglement and quantum tunnelling. 
    \item The correlation between the number of likes and the explanatory videos' explaining quality is moderate and statistically significant for the total sample ($r = 0.37$, $p < 0.01$) and for the sub-sample including quantum entanglement videos ($r = 0.42$, $p < 0.05$). For the videos on quantum tunnelling, however, the correlation is not statistically significant.
    \item The correlation between the number of dislikes and the explanatory videos' explaining quality is moderate and statistically significant for the total sample ($r = 0.32$, $p < 0.05$). In contrast, it is not statistically significant for the videos on quantum entanglement and quantum tunnelling.
\end{itemize}

Our results compare well with the findings reported earlier for the mechanics context~\parencite[cf.][]{Kulgemeyer2016neu}: While the correlations presented in both studies seem different at first glance (cf. table~\ref{tab:comp1}), we note that most of the correlations reported by~\textcite{Kulgemeyer2016neu} fall within the $95\%$ confidence intervals of our correlation coefficients (or vice versa).\\
In additon, our results also shed new light on the underlying relationships: In their 2016 article \textcite{Kulgemeyer2016neu} found no statistically significant correlation between the videos' explaining quality and the number of likes although the authors expected such a correlation due to the 'illusion of understanding': ``Students do not realise the possible inconsistencies in their understanding and feel as if they have understood a topic''~\parencite[][p. 11]{Kulgemeyer2016neu}. This assumption is supported by empirical evidence from a recently published experimental study by \textcite{Kulgemeyer2022}. For the explanatory videos on quantum topics included in our sample, we indeed uncovered a statistically significant correlation between the number of likes and the videos' explaining quality ($r = 0.37$, $p <0.01$).

{\linespread{1}
\begin{table}[]
\caption{ Pearson's correlation coefficient $r$ between the measure of explaining quality (in CP) and the surface features provided by YouTube. 
For the correlations calculated in our study, we report $95\%$ confidence intervals ($95\%$-CI).}
\label{tab:comp1}
\begin{tabular}{p{0.35\textwidth}p{0.25\textwidth}p{0.1\textwidth}p{0.2\textwidth}}
\toprule
                       & Surface feature & $r$ & $95\%$-CI   \\
\midrule
Explaining quality          & Time online (days) & $-0.14$ & $[-0.38; 0.12]$ \\
Videos on quantum topics~\tabfnm{a}    & Views              & $0.27^{*}$  & $[0.02; 0.49]$\\
     & Likes              & $0.37^{**}$ & $[0.13; 0.57]$\\
                            & Dislikes            & $0.32^{*}$  & $[0.06; 0.53]$\\
                            & Relevant comments  & $0.46^{**}$ & $[0.24; 0.64]$\\
\midrule
Explaining quality          & Time online (days) & $-0.05$      & - \\
Videos on mechanics topics~\tabfnm{b}  & Views              & $-0.26$      & - \\
\parencite[][]{Kulgemeyer2016neu}                 & Likes              & $0.21$       & -\\
     & Dislikes           & $-0.09$      & -\\
                            & Relevant comments  & $0.38^{**}$  & -\\
\bottomrule
\end{tabular}
{\linespread{1.5}
\begin{tablenotes}[para,flushleft]
        {\small
            \textit{Note.} Statistical significance of the correlations is denoted by an asterisk:
            \tabfnt{*}\textit{p} < .05.
            \tabfnt{**}\textit{p} < .01.
            \tabfnt{a} $N = 60$, this study.
            \tabfnt{b} $N = 51$, \parencite[cf.][p. 10]{Kulgemeyer2016neu}.
            
         }
    \end{tablenotes}}
\end{table}
}

Moreover, we find the number of dislikes ($r = 0.32$, $p < 0.05$) and the number of views ($r = 0.27$, $p < 0.05$) to have statistically significant correlations with the explaining quality of the videos on quantum entanglement and tunnelling. In contrast, \textcite{Kulgemeyer2016neu} have not found the corresponding correlations to be statistically significant for the videos on classical mechanics topics. The analysis of partial correlations, though, puts these differences between the two studies into perspective: We controlled the correlations between the videos' explaining quality and the surface features provided by YouTube for the number of subscribers to the channels by which the videos were published. As a result, the correlation between explaining quality and views ($r = 0.23$) loses its statistical significance. To describe this observation, we go along with~\textcite{Kulgemeyer2016neu} who state that ``the number of views is more influenced by [...] the popularity of the YouTube channel than the explaining quality'' (p. 5). Accordingly, the correlation between explaining quality and dislikes ($r = 0.26$) loses its statistical significance, though remaining moderate (cf. table~\ref{tab:partial_subscribe}).\\

Lastly, we newly introduced the number of interactions, i.e., the sum of likes and dislikes for a given YouTube explanatory video, into the analysis (cf. table~\ref{tab:interaction}). The number of interactions correlates statistically significantly with the explaining quality of the explanatory videos on entanglement and tunnelling: $r = 0.39$, $p < 0.01$. The partial correlation - when controlling for the number of subscribers of the channels by which the videos are published - of $r = 0.43$, $p < 0.01$ was even higher.
\clearpage

\subsection{Discussion of research question 2}
Compared to the metrics provided by YouTube alongside each video (e.g., the number of views), the number of relevant comments turned out to be most strongly correlated with the explaining quality of explanatory videos ($r = 0.46$, $p < 0.01$ for the total sample) on (a) quantum entanglement ($r = 0.59$, $p < 0.01$), and (b) quantum tunnelling ($r = 0.31$, $p < 0.1$). Similarly, \textcite[][p. 10]{Kulgemeyer2016neu} report a correlation of $r = 0.38$ ($p < 0.01$) between explaining quality and the number of relevant comments for videos on Newton's third law and Kepler's laws, respectively.\\

We controlled the correlations between the videos' explaining quality and the number of relevant comments for the videos' time online (in days). As a result, the partial correlation between explaining quality and number of relevant comments for the total sample increased ($r = 0.55$, $p < 0.001$). This result is comparable to the one reported for the mechanics context, where a partial correlation coefficient of $p = 0.40$, $p < 0.01$ was found~\parencite{Kulgemeyer2016neu}. \\

The medium to high correlation between the explanatory videos' explaining quality and the number of relevant comments might be justified via the users' cognitive activation: ``Hence, videos that accumulate plenty of those relevant comments are more successful in catching viewers’ attention as these videos might use either a more stimulating explanation or the explanation delivered is considered as a starting point for further learning progress''~\parencite[][p. 12]{Kulgemeyer2016neu}.

\section{Conclusion}
\label{sec7}
Our results support the findings presented earlier for YouTube explanatory videos on mechanics~\parencite[cf.][]{Kulgemeyer2016neu}, according to which 
\begin{itemize}
    \item there is a statistically significant correlation between explaining quality and the number of content-related comments ($r = 0.46$, $p < 0.001$ in our study, cf. table~\ref{tab:corr_overall}), and 
    \item YouTube's surface metrics (e.g., likes) might not be fruitful indicators for the explaining quality of explanatory videos (cf. table~\ref{tab:partial_subscribe}). 
\end{itemize}
However, focusing on YouTube explanatory videos addressing quantum entanglement and tunnelling, our study contributes to extending previous results presented by~\textcite{Kulgemeyer2016neu} in two respects: 
\begin{enumerate} 
    \item We find a statistically significant correlation between the number of likes and the explaining quality of explanatory videos on the quantum topics entanglement and tunnelling ($r = 0.37$, $p < 0.01$, cf. table~\ref{tab:corr_overall}). Although such a correlation has already been assumed in the previous study~\parencite[cf.][]{Kulgemeyer2016neu}, it could not be found at that time in the context of explanatory videos on topics of classical mechanics.
    \item Our study hints that the number of interactions (e.g., the sum of likes and dislikes) might be an indicator for videos of high explaining quality ($r = 0.39$, $p < 0.01$, cf. table~\ref{tab:interaction}). We argue that this result fits well to the number of relevant comments being statistically significantly correlated with the explaining quality of explanatory videos (cf. table~\ref{tab:corr_overall}).
\end{enumerate}

\subsection{Limitations}
It is important to note that the results presented in this article should be interpreted with caution for the following reasons: 
\begin{enumerate} 
    \item We could only include a small number of $N = 60$ videos in our sample due to the huge amount of data and the great effort required for data analysis (e.g., categorization of all comments underneath each video).
    \item Classical correlations, as presented in this article, allow for the exploration of relationships between variables, but not for the identification of causal connections. 
    \item The data analysis is largely based on the metrics provided by YouTube, which are not fully transparent to users (cf.~\cite{Kulgemeyer2016neu}). 
    \item In this study, we only analyzed explanatory videos on the topics quantum entanglement and tunnelling, and hence, the correlations found are not generalizable to different topics.
\end{enumerate}

\subsection{Outlook}
Despite the above-mentioned limitations, our results may serve as a valuable starting point for future research, in particular with respect to teaching and learning quantum concepts: While in this study only scientifically sound explanatory videos have been included for the analysis, the internet is crowded with scientifically misleading or mystifying explanatory videos on quantum concepts, such as quantum entanglement and quantum tunnelling. Since YouTube's surface features, however, are not likely to provide reliable quality indicators, future educational research should (a) explore widespread misconceptions in explanatory videos on quantum concepts, and (b) make further efforts towards the derivation of evidence-based selection criteria that support both students and teachers/lecturers in detecting high quality content out of the dark noise.\\

\subsection{Data availability:} 
The data presented in this study are available on request from the corresponding author.
\subsection{Disclosure Statement:} No potential conflict of interest was reported by the author(s). 
\subsection{Ethical statement:} The study met the ethics requirements of the University of Erlangen at the time the data was collected.

{\linespread{1}
\printbibliography
}

\section*{Appendix}

{\linespread{1}
\begin{table}[H]
\caption{The category system to evaluate explanatory videos' explaining quality~\parencite[][p. 7]{Kulgemeyer2016neu}. The sub-categories that are related to a decrease of explaining quality, and hence, lead to a negative point ($= -1$ CP) are marked with \mbox{'-'}. The categories used in this context are also integrated in the framework of effective explanation videos~\parencite[cf.][p. 2449]{Kulgemeyer2020Framework}. For an in-depth description of all the categories we refer the reader to~\textcite{Kulgemeyer2020Framework}.}
\label{tab:cat}
\resizebox{16.5cm}{!}{
\begin{tabular}{p{0.3\textwidth}p{0.63\textwidth}}
\toprule
\textbf{Main category} & \textbf{Subcategories} \\
\midrule
Content & 1. Scientific mistake (-) \newline 2. Mistake corrected \\
Structure & 3. Giving an outlook \newline 4. Giving a review \newline 5. Giving a summary \newline 6. Ignoring students' comment (-) \newline 7. Emphazising important points \newline 8. Open justification of the explaining approach \newline 9. Addressing common misconceptions \\
Use of language & 10. Paraphrasing technical terms \newline 11. Comment technical term with everyday language \newline 12. Comment technical term with other technical \phantom{abc} terms \newline 13. Leaving new technical term uncommented (-) \\
Contexts and examples & 14. Addressing explainee \newline 15. Example close to everyday life \newline 16. Abstract example \newline 17. Without context (-) \newline 18. Connecting at least two examples by showing \phantom{abc} analogies \newline 19. Connecting example to explained topic by showing analogies \\
Mathematics & 20.  Providing numerical example for formula \newline 21. Using formula \newline 22. Describing relationships by use of ‘the more... the less/more’ relations \newline 23. Using mathematical terms and idealisations  \\
Interrogation & 24. Asking further questions \\
Non-verbal elements & 25. Using realistic figures (such as photos) \newline 26. Using analogical figures \newline 27. Using logical figures (such as diagrams) \newline 28. Using experiments \newline 29. Connecting non-verbal elements \newline 30. Using writings \newline 31. Draw/amend figures \\
\bottomrule
\end{tabular}}

\end{table}}

\end{document}